\documentstyle[11pt]{article}
\newcommand{\reff}[1]{eq.~(\ref{#1})}

\begin{document}


\def\a{\alpha}
\def\b{\beta}
\def\d{\delta}
\def\e{\epsilon}
\def\g{\gamma}
\def\k{\kappa}
\def\l{\lambda}
\def\o{\omega}
\def\t{\theta}
\def\s{\sigma}
\def\D{\Delta}
\def\L{\Lambda}

\def\G{{\cal G}}
\def\C{{\bf C}}
\def\P{{\bf P}}

\def\uqoh{U_q(osp(2|2n)^{(1)})}
\def\uqo{U_q(osp(2|2n))}
\def\uqgh{U_q(sl(m|n)^{(1)})}
\def\uqg{U_q(sl(m|n))}
\def\oh{osp(2|2n)^{(1)}}
\def\gh{sl(m|n)^{(1)}}
\def\qh{q^{-h_0/2}}


\def\beq{\begin{equation}}
\def\eeq{\end{equation}}
\def\bea{\begin{eqnarray}}
\def\eea{\end{eqnarray}}
\def\ba{\begin{array}}
\def\ea{\end{array}}
\def\no{\nonumber}
\def\lt{\left}
\def\rt{\right}
\newcommand{\bq}{\begin{quote}}
\newcommand{\eq}{\end{quote}}

\newtheorem{Theorem}{Theorem}
\newtheorem{Definition}{Definition}
\newtheorem{Proposition}{Proposition}
\newtheorem{Lemma}[Theorem]{Lemma}
\newtheorem{Corollary}[Theorem]{Corollary}
\newcommand{\proof}[1]{{\bf Proof. }
        #1\begin{flushright}$\Box$\end{flushright}}

\begin{titlepage}
\begin{flushright}
BI-TP-94/33\\
UQMATH-94-05\\
hep-th/9408006
\end{flushright}
\vskip.3in
\begin{center}
{\huge On Type-I Quantum Affine Superalgebras}
\vskip.3in
{\Large Gustav W. Delius\footnote{Supported by Habilitationsstipendium der
Deutschen Forschungsgesellschaft}
\footnote{On leave from Department of Physics, Bielefeld University,
Germany}}
\vskip.1in
{\large Department of Mathematics, King's College London, Stand,
London WCR 2LS, UK}
email: delius@mth.kcl.ac.uk
\vskip.2in
{\Large Mark D. Gould, Jon R. Links, Yao-Zhong Zhang}
\vskip.1in
{\large Department of Mathematics, University of Queensland, Brisbane,
Qld 4072, Australia}
email: yzz@maths.uq.oz.au
\end{center}
\vskip.4in
\begin{center}
{\bf Abstract:}
\end{center}
\begin{quote}
The type-I simple Lie-superalgebras are $sl(m|n)$ and $osp(2|2n)$.
We study the quantum deformations of their untwisted affine extensions
$U_q(sl(m|n)^{(1)})$ and $U_q(osp(2|2n)^{(1)})$.

We identify additional relations between the simple generators (``extra
$q$-Serre relations") which need to be imposed to properly define
$\uqgh$ and $U_q(osp(2|2n)^{(1)})$.

We present a general technique for deriving the spectral parameter
dependent R-matrices from quantum affine superalgebras.

We determine the R-matrices for the type-I affine superalgebra
$U_q(sl(m|n)^{(1)})$ in
various representations, thereby deriving new solutions of the
spectral-dependent Yang-Baxter equation. In particular, because this
algebra possesses one-parameter families of finite-dimensional irreps,
we are able to construct R-matrices depending on two additional
spectral-like parameters, providing generalizations of the
free-fermion model.
\end{quote}

\end{titlepage}
\newpage

\newcommand{\sect}[1]{\setcounter{equation}{0}\section{#1}}
\renewcommand{\theequation}{\thesection.\arabic{equation}}

\sect{Introduction\label{intro}}

Quantum affine algebras are q-deformations
\cite{Dri85,Dri86,Jim85,Jim86a,Jim86b}
of the  enveloping algebras of affine Kac-Moody algebras \cite{Kac90}. They
were
introduced as a powerful tool for the construction of solutions of the
spectral parameter dependent Yang-Baxter equation. They are therefore
the algebraic structures underlying the integrable models of statistical
mechanics (commuting transfer matrices) as well as the quantum integrable
field theories (quantum inverse scattering method, factorizable
scattering matrices).

There are however solutions to the Yang-Baxter equation, and thus there
are integrable models, which do not come from a quantum affine bosonic algebra.
Examples are the Perk-Schultz model \cite{deV91,Mar92} and the free fermion
model \cite{Baz85}. It is now understood that the algebras underlying these
models are quantum affine superalgebras.
In spite of their significance, quantum affine superalgebras have so far
remained essentially unstudied in the literature. In this paper we will
study the type-I untwisted affine superalgebras
$U_q(sl(m|n)^{(1)})$ and $U_q(osp(2|2n)^{(1)})$.

Lie superalgebras are by no means just straightforward generalizations
of their bosonic counterparts. Rather they are much richer structures
and have a more complicated representation theory \cite{Kac77,Kac78}.
Two interesting effects occur when one defines the quantum deformations
of the enveloping algebras of Lie superalgebras. One is that a fixed
Lie superalgebra allows many inequivalent systems of simple roots and
that these give rise to different Hopf algebras upon deformation. They
are related by twistings \cite{Kho94}. The other is that the simple
raising and lowering generators of a Lie superalgebra obey more
relations than just the usual Serre relations known from bosonic Lie
algebras \cite{Sch92,Sch93,Kho91,Kho93,Yam91,Yam94}.
We give the necessary
extra Serre relations to properly define $\uqgh$ and
$U_q(osp(2|2n)^{(1)})$ in section \ref{secSerre}.

To every pair of finite-dimensional irreps of a quantum affine algebra
there exists a spectral parameter dependent R-matrix. It  can in
principle be constructed by inserting the representation matrices into
the general formula for the universal R-matrix. The universal R-matrices
are known also for quantum affine superalgebras \cite{Kho93}. In practice
however, this method is too difficult and therefore, in section
\ref{secmethod},
we present a practical method to obtain the R-matrices. Our method
immediately gives the spectral decomposition of the R-matrices, i.e.,
it gives the R-matrices in the form
\beq
\breve{R}(x)=\sum_\nu \rho_\nu(x) \P_\nu
\eeq
where the $\P_\nu$ are elementary intertwining operators of the
non-affine algebra and the $\rho_\nu(x)$ are meromorphic functions
of the spectral parameter $x$.
In section \ref{secexamples} we determine these functions
$\rho_\nu(x)$ explicitly for a large number of examples.

Type-I superalgebras are particularly interesting because they possess
one-parameter families of finite-dimensional irreps. The R-matrices
for a pair of such representations will then also depend on these extra
parameters. These parameters enter the Yang-Baxter equation in a similar
way as the spectral parameter, though in a non-additive form. An R-matrix
of this form, depending on three ``spectral'' parameters, has been known:
the R-matrix of the free fermion model \cite{Baz85}. This paper gives the
quantum group theoretic interpretation for this R-matrix (in its
trigonometric form): it is the R-matrix for the tensor product of two
parameter-dependent 2-dimensional irreps of $U_q(sl(1|1)^{(1)})$. This
allows us to give the generalizations of this R-matrix for other
type-I affine superalgebras in sections \ref{secpar1}.
We expect that these R-matrices will find interesting
physical applications.

This paper is organized as follows: In section \ref{secSerre} we give the
definition of the type-I untwisted quantum affine superalgebras
$U_q(sl(m|n)^{(1)})$ and $U_q(osp(2|2n)^{(1)})$. In particular we identify
extra Serre relations needed to define them.
In section \ref{secmethod} we give a general techique for deriving
the spectral parameter dependent R-matrices
from these quantum affine superalgebras. In section \ref{secexamples}
we apply this technique
to derive a large number of new R-matrices and in particular the novel
R-matrices depending on extra spectral-like parameters.
Section \ref{secdiscussion} contains a discussion of our results.
For the convenience of the reader, two appendices review some known
material on the representation theory of superalgebras. Appendix
\ref{appinduced} reviews Kac's induced module construction and
appendix \ref{appclass} gives the results about the classification
of finite-dimensional unitary irreps of type-I superalgebras.

\sect{Definition of Quantum Affine Superalgebras and
        extra Serre relations\label{affinization}\label{secSerre}}

In this section we want to define a quantum affine superalgebra
$U_q(\G^{(1)})$ as the
quantum deformation, depending on a nonzero parameter $q\in{\bf C}$,
of the universal enveloping algebra of an untwisted
simple affine superalgebra $\G^{(1)}$.

Let $\G$ be a simple superalgebra \cite{Kac77}.
Throughout the paper we choose the distinguished\footnote{
As mentioned in the introduction, (affine) superalgebras allow many
inequivalent systems of simple roots. See \cite{Kac77}. The relation
between the different quantum superalgebras obtained by choosing
different systems of simple roots is studied in \cite{Kho94}.}
set of simple roots $\a_i\,,~i=1,\cdots,r$.
Let $(~,~)$ be a fixed invariant bilinear form on the root space
of $\G$.
Let $\G^{(1)}$ denote the untwisted affine superalgebra
associated with $\G$.
It possesses an additional simple root
$\a_0=\d-\psi$, where $\d$ is the minimal
imaginary root of $\G^{(1)}$ satisfying $(\d,\d)=0=(\d,\a_i),~\forall i$,
and $\psi$ denotes the highest root of $\G$.
The generalized Cartan matrix $A=(a_{ij})_{0\leq i,j\leq r}$ is
defined from the simple roots by
\beq
a_{ij}=\left \{
\begin{array}{l}
\frac{2(\a_i,\a_j)}{(\a_i,\a_i)},~~~~{\rm if}~(\a_i,\a_i)\neq 0\\
(\a_i,\a_j),~~~~{\rm if}~(\a_i,\a_i)=0
\end{array}
\right .
\eeq

In order to be able to define the quantum deformation $U_q(\G^{(1)})$
we present $\G^{(1)}$ entirely in terms of its
generators $\{e_i,~f_i,~h_i,~i=0,1,...,r\}$ corresponding to the simple
roots. In analogy to the procedure used for bosonic Lie algebras we
define $\overline{\G^{(1)}}$ as the Lie superalgebra generated by the
simple generators $\{e_i,~f_i,~h_i,~i=0,1,...,r\}$
\footnote{Note that throughout this paper we define the affine algebras
without the derivation $d$. It can be reintroduced at any stage without
any complications.} subject to the relations
\bea\label{Lie}
&&[h_i,h_j]=0,~~~~~[e_i,f_j]=\delta_{ij}h_j,\nonumber\\
&&[h_i,e_j]=(\a_i,\a_j)\ e_j,~~~~~
[h_i,f_j]=-(\a_i,\a_j)\ f_j,\nonumber\\
&&[e_i,e_i]=[f_i,f_i]=0,~~~\mbox{if } (\a_i,\a_i)=0,\\
&&(ad\ e_i)^{1-a_{ij}}\ e_j=0,~~~
(ad\ f_i)^{1-a_{ij}}\ f_j=0,~~~\mbox{if } (\a_i,\a_i)\neq 0,~~i\neq j.
\nonumber
\eea
Here $[~,~]$ denotes the Lie superbracket wich satisfies
$[a,b]=-(-1)^{[a][b]}[b,a]$ where $[a]\in{\bf Z}_2$ denotes the degree of
the element $a$, and $(ad\ a)\ b\equiv [a,b]$. The last relations in
\reff{Lie} are the usual Serre relations.

The universal enveloping algebra $\overline{U(\G^{(1)})}$
of $\overline{\G^{(1)}}$ is the unital ${\bf Z}_2$-graded associative
algebra generated by the same generators and the same relations
\reff{Lie} where now however $[~,~]$ denotes the (anti)commutator
$[a,b]=a\,b-(-1)^{[a][b]}b\,a$. The Lie superalgebra $\overline{\G^{(1)}}$
is naturally embedded in $\overline{U(\G^{(1)})}$.

While in the purely bosonic case $\overline{\G^{(1)}}=\G^{(1)}$, this is
not generally true for Lie superalgebras. Rather, the Lie superalgebra
$\overline{\G^{(1)}}$ defined above generally contains a proper maximal
Lie algebra ideal $M$ and the affine superalgebra $\G^{(1)}$ is
obtained as the quotient $\overline{\G^{(1)}}/M$ \cite{Sch92,Sch93}.
The extra relations among the simple generators which hold in $\G^{(1)}$
are refered to as ``extra Serre relations''. The universal enveloping
algebra $U(\G^{(1)})$ is obtained as the quotient
$U(\G^{(1)})=\overline{U(\G^{(1)})}/\tilde{M}$ where $\tilde{M}$ is
the two-sided ideal in $\overline{U(\G^{(1)})}$ generated by $M$.

We now introduce the quantum deformation $\overline{U_q(\G^{(1)})}$ of the
universal enveloping algebra $\overline{U(\G^{(1)})}$ as the
unital ${\bf Z}_2$-graded associative
algebra generated by $\{e_i,~f_i,~q^{\pm h_i},~i=0,1,...,r\}$ subject to the
relations
\begin{eqnarray}
&&q^h.q^{h'}=q^{h+h'}\,,~~~~h,~ h'=\pm h_i,~ i=0,1,...,r\nonumber\\
&&q^{h_i}e_jq^{-h_i}=q_i^{a_{ij}} e_j\,,~~q^{h_i}f_jq^{-h_i}=q_i^{-a_{ij}}
  f_j\nonumber\\
&&[e_i, f_j]=\delta_{ij}\frac{q^{h_i}-q^{-h_i}}{q-q^{-1}}\nonumber\\
&&e_i^2=f_i^2=0\,,~~~~~~{\rm if}~(\a_i,\a_i)=0\no\\
&&\sum^{1-a_{ij}}_{k=0}(-1)^k \left [
\begin{array}{c}
1-a_{ij}\\
k
\end{array}
\right \}_{q_i} e_i^{1-a_{ij}-k}e_je_i^{k}
   =0,~~~(i\neq j),~~~(\a_i,\a_i)\neq 0\nonumber\\
&&\sum^{1-a_{ij}}_{k=0}(-1)^k \left [
\begin{array}{c}
1-a_{ij}\\
k
\end{array}
\right \}_{q_i} f_i^{1-a_{ij}-k}f_jf_i^{k}
   =0,~~~(i\neq j),~~~(\a_i,\a_i)\neq 0\label{defining-relations}
\end{eqnarray}
where
\begin{equation}
\left [
\begin{array}{c}
1-a_{ij}\\
k
\end{array}
\right \}_{q_i}=\left \{
\begin{array}{l}
\left [
\begin{array}{c}
1-a_{ij}\\
k
\end{array}
\right ]^-_{q_i}\,,~~~~{\rm if}~\a_i~{\rm is~even}\\
(-)^{k(k-(-1)^{[\a_j]})/2}\left [
\begin{array}{c}
1-a_{ij}\\
k
\end{array}
\right ]^+_{q_i}\,,~~~{\rm if}~\a_i~{\rm is~odd}
\end{array}
\right .
\end{equation}
with
\bea
&&\left [
\begin{array}{c}
m\\
n
\end{array}
\right ]^\pm_q=\left \{
\begin{array}{l}
   \frac{[m]^\pm !}{[m-n]^\pm ![n]^\pm !}, ~~~m\geq n\geq 0\\
0,~~~~~{\rm otherwise}
\end{array}
\right .\,,\no\\
&&[m]^\pm !=[m]^\pm [m-1]^\pm\cdots [0]^\pm\,,~~~~[m]^\pm=\frac{q^m\pm
  q^{-m}}{q-q^{-1}},~~~~[0]^\pm=1
\eea
\beq
q_i=\left \{
\begin{array}{ll}
q^{(\a_i,\a_i)/2}\,, & ~~{\rm if}~(\a_i,\a_i)\neq 0\\
q\,, & ~~{\rm if}~ (\a_i,\a_i)=0
\end{array}
\right .
\eeq
Throughout this
paper we will assume that $q$ is generic, i.e. not a root of unity.
In the limit $q\rightarrow 1$ the above relations reproduce the relations
\reff{Lie} and thus $\overline{U_q(\G^{(1)})}$ goes over to
$\overline{U(\G^{(1)})}$.

The algebra $\overline{U_q(\G^{(1)})}$ is a Hopf algebra.
The coproduct is given by
\begin{eqnarray}\label{coproduct}
&&\Delta(q^{\pm h})=q^{\pm h}\otimes q^{\pm h}\nonumber\\
&&\Delta(e_i)=e_i\otimes q^{-h_i/2}+q^{h_i/2}\otimes e_i\\
&&\Delta(f_i)=f_i\otimes q^{-h_i/2}+q^{h_i/2}\otimes f_i\nonumber
\end{eqnarray}
We omit the formulas for the antipode and the counit.

Finally we are ready to define the quantum affine superalgebra
$U_q(\G^{(1)})$ as the quotient
$U_q(\G^{(1)})=\overline{U_q(\G^{(1)})}/\tilde{M}_q$,
where $\tilde{M}_q$ is a proper two-sided Hopf algebra ideal in
$\overline{U_q(\G^{(1)})}$ which, in the limit $q\rightarrow 1$,
goes over to the ideal $\tilde{M}$ in $\overline{U(\G^{(1)})}$.
This ensures that $U_q(\G^{(1)})$ is a Hopf algebra which goes over
to the undeformed algebra $U(\G^{(1)})$ in the limit of $q\rightarrow 1$.
The extra relations among the simple generators which arise from the
division by the ideal $\tilde{M}_q$ are refered to as ``extra $q$-Serre
relations". In the following we will identify the
extra $q$-Serre relations for $\uqgh$ and $\uqoh$.

All of the above definitions can also be taken over to the non-affine
case with the only change that all indices run only from $1$ to $r$, i.e.,
there are no $e_0, f_0$ and $h_0$.

\subsection{Extra $q$-Serre relations for $\uqgh$\label{secgl}}

Suitable extra q-Serre relations for the case
$\G=sl(m|n)$ have been derived in
\cite{Sch92,Sch93}, see also note added to \cite{Kho91}. These can
be straightforwardly extended to $\G^{(1)}=sl(m|n)^{(1)}$.

For $sl(m|n)^{(1)}$ we take the set of simple roots,
\bea
&&\a_0=\d-\e_1+\d_n\,,\no\\
&&\a_i=\e_i-\e_{i+1}\,,~~~~~i=1,2,\cdots, m-1\no\\
&&\a_m=\e_m-\d_1\,,\no\\
&&\a_{m+j}=\d_j-\d_{j+1}\,,~~~~~j=1, 2,\cdots,n-1
\eea
with $\d,~\{\e_i\}_{i=1}^m$ and $\{\d_i\}_{i=1}^n$ satisfying
\bea
&&(\d,\d)=(\d,\e_i)=(\d,\d_i)=0,~~~~(\e_i,\e_j)=\d_{ij},\no\\
&&(\d_i,\d_j)=-\d_{ij},~~~~(\e_i,\d_j)=0
\eea
The associated Dynkin diagram is
\beq
\unitlength=1mm
\linethickness{0.4pt}
\begin{picture}(126.00,32.00)(20,5)
\put(40.00,10.00){\circle{4.00}}
\put(52.00,10.00){\circle{4.00}}
\put(70.00,10.00){\circle{4.00}}
\put(94.00,10.00){\circle{4.00}}
\put(112.00,10.00){\circle{4.00}}
\put(124.00,10.00){\circle{4.00}}
\put(42.00,10.00){\line(1,0){8.00}}
\put(54.00,10.00){\line(1,0){4.00}}
\put(65.00,10.00){\line(1,0){3.00}}
\put(72.00,10.00){\line(1,0){8.00}}
\put(84.00,10.00){\line(1,0){8.00}}
\put(96.00,10.00){\line(1,0){4.00}}
\put(106.00,10.00){\line(1,0){4.00}}
\put(114.00,10.00){\line(1,0){8.00}}
\put(40.00,12.00){\line(5,2){40.00}}
\put(84.00,28.00){\line(5,-2){40.00}}
\put(82.00,26.00){\makebox(0,0)[ct]{$1$}}
\put(82.00,32.00){\makebox(0,0)[cb]{$\a_0$}}
\put(82.00,29.00){\makebox(0,0)[cc]{$\bigotimes$}}
\put(82.00,10.00){\makebox(0,0)[cc]{$\bigotimes$}}
\put(52.00,13.00){\makebox(0,0)[cb]{$1$}}
\put(70.00,13.00){\makebox(0,0)[cb]{$1$}}
\put(82.00,13.00){\makebox(0,0)[cb]{$1$}}
\put(94.00,13.00){\makebox(0,0)[cb]{$1$}}
\put(112.00,13.00){\makebox(0,0)[cb]{$1$}}
\put(125.00,13.00){\makebox(0,0)[cb]{$1$}}
\put(39.00,13.00){\makebox(0,0)[cb]{$1$}}
\put(40.00,7.00){\makebox(0,0)[ct]{$\a_1$}}
\put(52.00,7.00){\makebox(0,0)[ct]{$\a_2$}}
\put(70.00,7.00){\makebox(0,0)[ct]{$\a_{m-1}$}}
\put(82.00,7.00){\makebox(0,0)[ct]{$\a_m$}}
\put(94.00,7.00){\makebox(0,0)[ct]{$\a_{m+1}$}}
\put(112.00,7.00){\makebox(0,0)[ct]{$\a_{m+n-2}$}}
\put(126.00,7.00){\makebox(0,0)[ct]{$\a_{m+n-1}$}}
\end{picture}
\eeq
where the grey nodes $\bigotimes$ correspond to the fermionic simple roots.

In addition to the extra $q$-Serre relations given in \cite{Sch92,Sch93}
for $\uqg$,
\bea\label{e-S1}
&&e_me_{m-1}e_me_{m+1}+e_{m-1}e_me_{m+1}e_m+e_me_{m+1}e_me_{m-1}+\no\\
&&~~~~~~~~~~~~~~~~~~~~~+e_{m+1}e_me_{m-1}e_m
  -(q+q^{-1})e_me_{m-1}e_{m+1}e_m=0\,,\no\\
&&f_mf_{m-1}f_mf_{m+1}+f_{m-1}f_mf_{m+1}f_m+f_mf_{m+1}f_mf_{m-1}+\no\\
&&~~~~~~~~~~~~~~~~~~~~~+f_{m+1}f_mf_{m-1}f_m
  -(q+q^{-1})f_mf_{m-1}f_{m+1}f_m=0\,,
\eea
one has, for $\uqgh$, the following extra $q$-Serre relations involving $e_0$
and $f_0$
\bea\label{e-S2}
&&e_0e_1e_0e_{m+n-1}+e_1e_0e_{m+n-1}e_0+e_0e_{m+n-1}e_0e_1+\no\\
&&~~~~~~~~~~~~~~~~~~~~~+e_{m+n-1}e_0e_1e_0
  -(q+q^{-1})e_0e_1e_{m+n-1}e_0=0\,,\no\\
&&f_0f_1f_0f_{m+n-1}+f_1f_0f_{m+n-1}f_0+f_0f_{m+n-1}f_0f_1+\no\\
&&~~~~~~~~~~~~~~~~~~~~~+f_{m+n-1}f_0f_1f_0
  -(q+q^{-1})f_0f_1f_{m+n-1}f_0=0\,.
\eea

\begin{Definition}{\label{uqgh}}
We set $\uqgh$ to be the Hopf superalgebra generated by the defining relations
(\ref{defining-relations}) of $\overline{\uqgh}$ subject to the {\em extra}
$q$-Serre relations (\ref{e-S1}) and (\ref{e-S2}).
\end{Definition}

\subsection{Extra $q$-Serre relations for $\uqoh$}

The set of simple roots of $\oh$ can be expressed as
\bea
&&\a_0=\d-\e-\d_1,\no\\
&&\a_1=\e-\d_1,\no\\
&&\a_i=\d_{i-1}-\d_i,~~~~i=2,3,\cdots,n\no\\
&&\a_{n+1}=2\d_n
\eea
with $\d,~\e$ and $\{\d_i\}_{i=1}^n$ satisfying
\bea
&&(\d,\d)=(\d,\e)=(\d,\d_i)=0,~~~~(\e,\e)=-1,\no\\
&&(\d_i,\d_j)=\d_{ij},~~~~(\e,\d_i)=0\,.
\eea
The Dynkin diagram for $\oh$ is
\beq
\unitlength=1mm
\linethickness{0.4pt}
\begin{picture}(84.00,28.00)(20,8)
\put(40.00,20.00){\circle{4.00}}
\put(70.00,20.00){\circle{4.00}}
\put(82.00,20.00){\circle{4.00}}
\put(72.00,21.00){\line(1,0){8.00}}
\put(72.00,19.00){\line(1,0){8.00}}
\put(68.00,20.00){\line(-1,0){4.00}}
\put(42.00,20.00){\line(1,0){5.00}}
\put(39.00,18.00){\line(-5,-2){9.00}}
\put(39.00,22.00){\line(-5,2){9.00}}
\put(54.00,20.00){\makebox(0,0)[cc]{$\cdots$}}
\put(70.00,23.00){\makebox(0,0)[cb]{$2$}}
\put(82.00,23.00){\makebox(0,0)[cb]{$1$}}
\put(82.00,17.00){\makebox(0,0)[ct]{$\a_{n+1}$}}
\put(70.00,17.00){\makebox(0,0)[ct]{$\a_n$}}
\put(42.00,17.00){\makebox(0,0)[lt]{$\a_2$}}
\put(42.00,23.00){\makebox(0,0)[lb]{$2$}}
\put(31.00,28.00){\makebox(0,0)[lb]{$1$}}
\put(31.00,12.00){\makebox(0,0)[lt]{$1$}}
\put(25.00,26.00){\makebox(0,0)[rc]{$\a_0$}}
\put(25.00,14.00){\makebox(0,0)[rc]{$\a_1$}}
\put(28.00,14.00){\makebox(0,0)[cc]{$\bigotimes$}}
\put(28.00,26.00){\makebox(0,0)[cc]{$\bigotimes$}}
\put(73.00,20.00){\line(1,1){4.00}}
\put(73.00,20.00){\line(1,-1){4.00}}
\put(29.00,24.00){\line(0,-1){8.00}}
\put(27.00,24.00){\line(0,-1){8.00}}
\end{picture}
\label{dg}
\eeq

{}From the Dynkin diagram we immediately see that $\overline{\uqo}$ is
a subalgebra of $\overline{\uqoh}$. Furthermore, there is no extra
$q$-Serre relations for $\uqo$; namely, $\uqo\equiv\overline{\uqo}$.
However, for $\uqoh$, we have found the following extra $q$-Serre relations:
\bea
X_q&\equiv &e_2e_1e_0+e_1e_0e_2-e_2e_0e_1-e_0e_1e_2+(q+q^{-1})(e_0e_2e_1-
    e_1e_2e_0)=0,\no\\
Y_q&\equiv &f_2f_1f_0+f_1f_0f_2-f_2f_0f_1-f_0f_1f_2+(q+q^{-1})(f_0f_2f_1-
    f_1f_2f_0)=0.\label{extra-S}
\eea
Direct calculation shows that $X_q,\ Y_q$ generate a proper Hopf algebra
ideal $\tilde{M}_q$.

\begin{Definition}{\label{uqoh}}
We  set $\uqoh$ to be the Hopf superalgebra generated by the defining relations
(\ref{defining-relations}) of $\overline{\uqoh}$ subject to the {\em extra}
$q$-Serre relations (\ref{extra-S}).
\end{Definition}

Note that the extra $q$-Serre relations (\ref{e-S1}), (\ref{e-S2}) and
(\ref{extra-S}) have evident classical counterparts: in the $q\rightarrow 1$
limit they reduce to the extra Serre relations which need to be imposed to
properly define the corresponding classical affine superalgebras. These extra
Serre relations have also been found by Yamane \cite{Yam94}

\subsection{The universal R-matrix}

The algebras $\uqgh$ and $\uqoh$ are quasitriangular graded Hopf algebras,
which means the following:

Let $\D'$ be the opposite coproduct: $\D'=T\D$, where $T$ is the graded twist
map: $T(a\otimes b)=(-1)^{[a][b]}b\otimes a\,,~\forall a,b\in U_q(\G^{(1)})$.
Then $\Delta$ and $\Delta'$ are related by the universal R-matrix $R$
in $U_q(\G^{(1)})\otimes U_q(\G^{(1)})$ satisfying, among others,
the relations
\begin{eqnarray}
&&R\D(a)=\D'(a)R\,,~~~~~\forall a\in U_q(\G^{(1)})\label{intw}\\
&&(I\otimes \D)R=R_{13}R_{12}\,,~~~~(\D\otimes I)R=R_{13}R_{23}\label{hopf}
\end{eqnarray}
where if $R=\sum a_i\otimes b_i$ then $R_{12}=\sum a_i\otimes b_i\otimes 1$
, $R_{13}=\sum a_i\otimes 1\otimes b_i$ etc. It follows from (\ref{hopf})
that $R$ satisfies the QYBE
\beq
R_{12}R_{13}R_{23}=R_{23}R_{13}R_{12}.\label{uqybe}
\eeq
The multiplication rule for the tensor product is defined for elements
$a,b,c,d\in U_q(\G^{(1)})$ by
\begin{equation}\label{gradprod}
(a\otimes b)(c\otimes d)=(-1)^{[b][c]}(ac\otimes bd)
\end{equation}
The existence of the universal R-matrix follows from Drinfeld's double
construction \cite{Dri85,Dri86} provided the algebra has a
triangular decomposition. This triangular decomposition follows from
Theorem 1 in \cite{Sch93}.

For any $x\in{\C}^\times$, we define an automorphism $D_x$ of
$U_q(\G^{(1)})$ as
\beq
D_x(e_i)=x^{\delta_{i0}}e_i\,,~~~~D_x(f_i)=x^{-\delta_{i0}}f_i,
{}~~~~D_x(h_i)=h_i.
\eeq
The parameter $x$ is called the spectral parameter.
One obtains a
spectral parameter dependent universal R-matrix $R(x)$ in
$U_q(\G^{(1)})\otimes U_q(\G^{(1)})$ by setting
\beq\label{spectral-R}
R(x)=(D_x\otimes I)(R).
\eeq
It solves the spectral parameter dependent Yang-Baxter equation
\beq
R_{12}(x)R_{13}(xy)R_{23}(y)=R_{23}(y)R_{13}(xy)R_{12}(x).\label{qybe}
\eeq

\sect{General Solution of Jimbo's Equations for the R-matrices
\label{general}\label{secmethod}}

In this section we will present our general techique for determining
the spectral parameter dependent R-matrices for quantum superalgebras.
{}From now on, with an abuse of notation, we will set
\beq
e_0\equiv f_\psi\,,~~~~f_0\equiv e_\psi\,,~~~~h_0\equiv -h_\psi
\eeq

Let $\pi_\l,~\pi_\mu$ and $\pi_\nu$ be three irreps of a quantum
superalgebra $U_q(\G)$, afforded by
the irreducible modules $V(\l),~V(\mu)$ and $V(\nu)$ with highest weights $\l,
{}~\mu$ and $\nu$, respectively.
Assume that all $\pi_\l\,,~\pi_\mu$ and $\pi_\nu$ are
affinizable, i.e. they can be extended to finite dimensional irreps of the
corresponding quantum affine superalgebra $U_q(\G^{(1)})$. Let
\beq
R^{\l\mu}(x)=(\pi_{\l}\otimes \pi_{\mu})(R(x))
\eeq
where $x\in {\bf C}$ is the spectral parameter introduced in \reff{spectral-R}.
Then $R^{\l\mu}(x)$ satisfies the system of linear
equations \cite{Jim86b} deduced from the intertwining property (\ref{intw})
\begin{eqnarray}
&&R^{\l\mu}(x)\Delta^{\l\mu}(a)=\Delta'^{\l\mu}
  (a)R^{\l\mu}(x)\,,~~~\forall a\in U_q(\G),\nonumber\\
&&R^{\l\mu}(x)\left (x\pi_{\l}(e_0)\otimes \pi_{\mu}(q^{-h_0/2})+
  \pi_{\l}(q^{h_0/2})\otimes \pi_{\mu}(e_0)\right )\nonumber\\
&&~~~~~~  =\left (x\pi_{\l}(e_0)\otimes \pi_{\mu}(q^{h_0/2})
  +\pi_{\l}(q^{-h_0/2})\otimes \pi_{\mu}(e_0)\right )R^{\l\mu}
  (x)\label{r(x)1}
\end{eqnarray}
and satisfies the QYBE
in the tensor product module $V(\l)\otimes V(\mu)\otimes V(\nu)$
of three irreps:
\begin{equation}
R^{\l\mu}_{12}(x)R^{\l\nu}_{13}(xy)R^{\mu\nu}_{23}(y)
  =R^{\mu\nu}_{23}(y)R^{\l\nu}_{13}(xy)R^{\l\mu}_{12}(x).
\end{equation}
In the above,
$\Delta^{\l\mu}(a)\equiv(\pi_{\l}\otimes \pi_{\mu})\Delta(a)$.
The solution $R^{\l\mu}(x)$ to the above equations intertwines the
coproduct and opposite
coproduct of $U_q(\G^{(1)})$ in the representation
$(\pi_{\l}\otimes\pi_{\mu})({\rm ev}_x\otimes {\rm ev}_1)\Delta$.
Because this representation is irreducible for generic
$x$, by Schur's lemma the solution is uniquely determined up to a scalar
factor.
We fix this factor in such a way that
\begin{equation}
\breve{R}^{\l\mu}(x)\breve{R}^{\mu\l}(x^{-1})=I\label{unitary},
\end{equation}
which is usually called the unitarity condition in the literature.
The multiplicative
spectral parameter $x$ can be transformed into an additive spectral
parameter $u$ by $x=\mbox{exp}(u)$.

In all our equations we implicitly use the ``graded" multiplication rule of
\reff{gradprod}. Thus the $R$-matrix of a quantum superalgebra satisfies
a ``graded" Yang-Baxter equation which, when written as an ordinary
matrix equation, contains extra signs:
\begin{eqnarray}
&&\left(R^{\l\mu}(x)\right)_{\a\b}^{\a'\b'}
\left(R^{\l\nu}(xy)\right)_{\a'\g}^{\a''\g'}
\left(R^{\mu\nu}(y)\right)_{\b'\g'}^{\b''\g''}
(-1)^{[\a][\b]+[\g][\a']+[\g'][\b']}\nonumber\\
&&~~~~~=\left(R^{\mu\nu}(y)\right)_{\b\g}^{\b'\g'}
\left(R^{\l\nu}(xy)\right)_{\a\g'}^{\a'\g''}
\left(R^{\l\mu}(x)\right)_{\a'\b'}^{\a''\b''}
(-1)^{[\b][\g]+[\g'][\a]+[\b'][\a']},
\end{eqnarray}
where $[\alpha]$ denotes the degree of the basis vector $v_\alpha$.
However after a redefinition
\beq\label{redef}
\left(\tilde{R}^{\l\mu}\right)_{\a\b}^{\a'\b'}
=\left(R^{\l\mu}\right)_{\a\b}^{\a'\b'}\,
(-1)^{[\a][\b]}
\eeq
the signs disappear from the equation. Thus any solution of the ``graded"
Yang-Baxter equation arising from the R-matrix of a quantum superalgebra
provides also a solution of the standard Yang-Baxter equation after
the redefinition in \reff{redef}. The graded Yang-Baxter equation was
studied already in \cite{Kul82}.

Now introduce the graded permutation operator $P^{\l\mu}$
on the tensor product
module $V(\l)\otimes V(\mu)$ such that
\begin{equation}\label{gradperm}
P^{\l\mu}(v_\alpha\otimes v_\beta)=(-1)^{[\alpha][\beta]}
  v_\beta\otimes v_\alpha\,,~~
  \forall v_\alpha\in V(\l)\,,~v_\beta\in V(\mu)
\end{equation}
and set
\begin{equation}
\breve{R}^{\l\mu}(x)=P^{\l\mu}R^{\l\mu}(x).
\end{equation}
Then (\ref{r(x)1}) can be rewritten as
\begin{eqnarray}
&&\breve{R}^{\l\mu}(x)\Delta^{\l\mu}(a)=\Delta^{\mu\l}(a)
  \breve{R}^{\l\mu}(x)\,,~~~\forall a\in U_q(\G),\nonumber\\
&&\breve{R}^{\l\mu}(x)\left (x\pi_{\l}(e_0)\otimes\pi_{\mu}
  (q^{-h_0/2})+\pi_{\l}(q^{h_0/2})\otimes
  \pi_{\mu}(e_0)\right )\nonumber\\
&&~~~~~~  =\left (\pi_{\mu}(e_0)\otimes \pi_{\l}(q^{-h_0/2})+
  x\pi_{\mu}(q^{h_0/2})\otimes \pi_{\l}(e_0)\right )
  \breve{R}^{\l\mu}(x)\label{r(x)2}
\end{eqnarray}
and in terms of $\breve{R}^{\l\mu}(x)$ the QYBE becomes
\begin{equation}
(I\otimes\breve{R}^{\l\mu}(x))(\breve{R}^{\l\nu}(xy)
  \otimes I)(I\otimes\breve{R}^{\mu\nu}(y))=
  (\breve{R}^{\mu\nu}(y)\otimes I)(I\otimes \breve{R}^{\l\nu}(xy))
  (\breve{R}^{\l\mu}(x)\otimes I)
\end{equation}
both sides of which act from $V(\l)\otimes V(\mu)\otimes V(\nu)$
to $V(\nu)\otimes V(\mu)\otimes V(\l)$. Note that this equation,
if written in matrix form, does not have extra signs in the superalgebra
case. This is because the definition of the graded permutation operator
in \reff{gradperm} includes the signs of \reff{redef}.

Consider three special cases:
$x=0,~x=\infty$ and $x=1$. For these special values of $x$,
$\breve{R}^{\l\mu}(x)$ satisfies the spectral-free QYBE,
\begin{equation}
(I\otimes\breve{R}^{\l\mu})(\breve{R}^{\l\nu}
  \otimes I)(I\otimes\breve{R}^{\mu\nu})=
  (\breve{R}^{\mu\nu}\otimes I)(I\otimes \breve{R}^{\l\nu})
  (\breve{R}^{\l\mu}\otimes I).
\end{equation}
Moreover, from (\ref{r(x)2}), we have respectively, for $x=0$,
\begin{eqnarray}
&&\breve{R}^{\l\mu}(0)\Delta^{\l\mu}(a)=\Delta^{\mu\l}(a)
  \breve{R}^{\l\mu}(0)\,,~~~\forall a\in U_q(\G),\nonumber\\
&&\breve{R}^{\l\mu}(0)\left (
  \pi_{\l}(q^{h_0/2})\otimes\pi_{\mu}(e_0)\right )
  =\left (\pi_{\mu}(e_0)\otimes \pi_{\l}(q^{-h_0/2})\right )
  \breve{R}^{\l\mu}(0)\label{r(0)}
\end{eqnarray}
for $x=\infty$,
\begin{eqnarray}
&&\breve{R}^{\l\mu}(\infty)\Delta^{\l\mu}(a)=
  \Delta^{\mu\l}(a)
  \breve{R}^{\l\mu}(\infty)\,,~~~\forall a\in U_q(\G),\nonumber\\
&&\breve{R}^{\l\mu}(\infty)\left (\pi_{\l}(e_0)\otimes\pi_{\mu}
  (q^{-h_0/2})\right )=\left (\pi_{\mu}(q^{h_0/2})\otimes
  \pi_{\l}(e_0)\right )
  \breve{R}^{\l\mu}(\infty)\label{r(infty)}
\end{eqnarray}
and for $x=1$,
\begin{eqnarray}
&&\breve{R}^{\l\mu}(1)\Delta^{\l\mu}(a)=\Delta^{\mu\l}(a)
  \breve{R}^{\l\mu}(1)\,,~~~\forall a\in U_q(\G),\nonumber\\
&&\breve{R}^{\l\mu}(1)\left (\pi_{\l}(e_0)\otimes\pi_{\mu}
  (q^{-h_0/2})+\pi_{\l}(q^{h_0/2})\otimes
  \pi_{\mu}(e_0)\right )\nonumber\\
&&~~~~~~  =\left (\pi_{\mu}(e_0)\otimes \pi_{\l}(q^{-h_0/2})+
  \pi_{\mu}(q^{h_0/2})\otimes \pi_{\l}(e_0)\right )
  \breve{R}^{\l\mu}(1)\label{r(1)}.
\end{eqnarray}

Eqs.(\ref{r(0)}), (\ref{r(infty)}) and (\ref{r(1)}) respectively admit a
unique solution for any given two irreps of
$U_q(\G)$. We consider only the case where the
tensor product decomposition
\begin{equation}
V(\l)\otimes V(\mu)=\bigoplus_\nu V(\nu),\label{free}
\end{equation}
is multiplicity-free.

We first focus on the $x=1$ case, i.e. (\ref{r(1)}). In this case we have,
from (\ref{unitary}),
\beq
\breve{R}^{\l\mu}(1)\breve{R}^{\mu\l}(1)=I.
\eeq
Let ${\cal P}^{\l\mu}_\nu:~V(\l)\otimes V(\mu)\rightarrow V(\nu)$ be the
projection operators which satisfy
\beq
{\cal P}^{\l\mu}_\nu\;{\cal P}^{\l\mu}_{\nu'}=\delta_{\nu\nu'}
  {\cal P}^{\l\mu}_\nu\,,~~~~\sum_{\nu}{\cal P}^{\l\mu}_\nu=I.\label{project1}
\eeq
Following \cite{Del94b}, we define operators ${\bf P}^{\l\mu}_\nu$ by
\beq
{\bf P}^{\l\mu}_\nu
={\cal P}^{\mu\l}_\nu\;\breve{R}^{\l\mu}(1)=\breve{R}^{\l\mu}(1)
  \;{\cal P}^{\l\mu}_\nu.
\eeq
Then, by definition
\beq
\breve{R}^{\l\mu}(1)=\sum_{\nu}{\bf P}^{\l\mu}_\nu.
  \label{solution-1}
\eeq
It is easy to show that
\beq
{\bf P}^{\l\mu}_\nu\;{\cal P}^{\l\mu}_{\nu'}={\cal P}^{\mu\l}_{\nu'}\;
  {\bf P}^{\l\mu}_\nu
  =\delta_{\nu\nu'}{\bf P}^{\l\mu}_\nu,\label{projection2}
\eeq
and
\bea
{\bf P}^{\mu\l}_\nu\;{\bf P}^{\l\mu}_{\nu'}&=&{\cal P}^{\l\mu}_\nu
  \breve{R}^{\mu\l}(1)\breve{R}^{\l\mu}(1){\cal P}^{\l\mu}_{\nu'}\nonumber\\
  &=&{\cal P}^{\l\mu}_\nu{\cal P}^{\l\mu}_{\nu'}=\delta_{\nu\nu'}
  {\cal P}^{\l\mu}_\nu\label{projection3}
\eea
Eqs.(\ref{projection2},~\ref{projection3}) imply that the operators
${\bf P}^{\l\mu}_\nu$ are ``projection" operators.
As can be seen from (\ref{r(1)}), the ``projectors"
${\bf P}^{\l\mu}_\nu$ are intertwiners of $U_q(\G)$.
The general solution satisfying the first equation of (\ref{r(x)2}) thus is
a sum of these elementary intertwiners
\beq
\breve{R}^{\l\mu}(x)=\sum_{\nu}\rho_\nu(x)\;
  {\bf P}^{\l\mu}_\nu\label{r-general}
\eeq
where $\rho_\nu(x)$ are some functions of $x$.
Our task is now to determine these functions so that (\ref{r-general})
satisfies also the second equation of (\ref{r(x)2}).

We begin by determining $\rho_\nu(x)$ at the special value $x=0$.
At $x=0$ the R-matrix $\breve{R}^{\l\mu}(0)$
reduces to the R-matrix of the quantum superalgebra
$U_q({\G})$ in $V(\l)\otimes V(\mu)$:
$\breve{R}^{\l\mu}(0)\equiv \breve{R}^{\l\mu}$.
We recall the following fact about the universal R-matrix $\bar{R}$ for
$U_q(\G)$:
\beq
\bar{R}^T\bar{R}=(v\otimes v)\D(v^{-1})\,,~~~v=uq^{-2h_\rho}\,,~~~
  \pi_\l(v)=q^{-C(\l)}\cdot I_{V(\l)}
\eeq
with $u=\sum_i\,(-1)^{[i]}\,S(\bar{b}_i)\bar{a}_i$ where
$\bar{R}=\sum_i\bar{a}_i\otimes\bar{b}_i$. $S$ is the antipode for $U_q(\G)$,
$C(\l)=(\l,\l+2\rho)$
is the eigenvalue of the quadratic Casimir element of $\G$ in an irrep with
highest weight $\l$;~ $\rho$ is the graded half-sum of positive roots of $\G$.
We now compute $\breve{R}^{\mu\l}\breve{R}^{\l\mu}$. Using the above
equations we have
\bea
\sum_\nu(\rho_\nu(0))^2{\cal P}^{\l\mu}_\nu&=&
  \breve{R}^{\mu\l}\breve{R}^{\l\mu}\equiv
  P^{\mu\l}\bar{R}^{\mu\l}P^{\l\mu}\bar{R}^{\l\mu}\nonumber\\
&=&(\bar{R}^T)^{\l\mu}\bar{R}^{\l\mu}=(\pi_\l\otimes\pi_\mu)(\bar{R}^T
  \bar{R})\nonumber\\
&=&\pi_\l(v)\otimes\pi_\mu(v)(\pi_\l\otimes\pi_\mu)\D(v^{-1})\nonumber\\
&=&\sum_{\nu} q^{C(\nu)-C(\l)-C(\mu)}{\cal P}^{\l\mu}_\nu
\eea
where use has been made of
$\breve{R}^{\l\mu}\equiv P^{\l\mu}\bar{R}^{\l\mu}
\equiv P^{\l\mu}(\pi_\l\otimes \pi_\mu)(\bar{R})$. It follows immediately that
\beq\label{rho0}
\rho_\nu(0)=\e(\nu)q^{\frac{C(\nu)-C(\l)-C(\mu)}{2}}
\eeq
where $\e(\nu)$ is the {\it parity} of $V(\nu)$ in $V(\l)\otimes V(\mu)$: it
arises as an eigenvalue of the unitary self-adjoint operator $R^{\l\mu}(1)$
at $q=1$ (see \cite{Del94b} for an explanation). Thus
\beq
\breve{R}^{\l\mu}(0)=\sum_{\nu}\e(\nu)
q^{\frac{ C(\nu)-C(\l)-C(\mu)}{2}}
  {\bf P}^{\l\mu}_\nu.\label{solution-2}
\eeq
A similar relation for quantum bosonic algebras was obtained in
\cite{Res88,Dri90}.

The $\rho_\nu(x)$ at the special value $x=\infty$ can be easily obtained from
(\ref{solution-2}) with the help of the unitarity condition (\ref{unitary}):
\beq
\breve{R}^{\l\mu}(\infty)=\sum_{\nu}
  \epsilon(\nu)\; q^{-\frac{C(\nu)-C(\l)-C(\mu)}{2}}\;
  {\bf P}^{\l\mu}_\nu.\label{solution-3}
\eeq

\noindent {\bf Remark:} both ${\cal P}^{\l\mu}_\nu$ and ${\bf P}^{\l\mu}_\nu$
can be determined by using pure representation theory of $U_q(\G)$ as follows.
Let $\{|e^\nu_\alpha\rangle_{\l\otimes \mu}\}$ be an orthonormal
basis for $V(\nu)$ in $V(\l)\otimes V(\mu)$.
$V(\nu)$ is also embedded in $V(\mu)\otimes V(\l)$ through the
opposite coproduct $\Delta'$. Let
$\{|e^\nu_\alpha\rangle_{\mu\otimes \l}\}$ be the
corresponding orthonormal basis. Then for real generic $q>0$,
${\cal P}^{\l\mu}_{\nu}$
and ${\bf P}^{\l\mu}_{\nu}$ may be written as
\begin{eqnarray}
&&{\cal P}^{\l\mu}_\nu=\sum_\alpha \left |e^\nu_\alpha\right
  \rangle_{\l\otimes
  \mu~\l\otimes\mu}\!\left\langle e^\nu_\alpha\right |,\nonumber\\
&&{\bf P}^{\l\mu}_\nu=\sum_\alpha\left |e^\nu_\alpha\right
  \rangle_{\mu\otimes\l~
  \l\otimes\mu}\!\left\langle e^\nu_\alpha\right |\,.
\end{eqnarray}
which should extend to all complex $q$ via analytic continuation arguments.
For more details see \cite{Del94c}.

Now we have sufficient relations to solve (\ref{r(x)2}) generally.
Inserting (\ref{r-general}) into the second
equation of (\ref{r(x)2}), and using (\ref{r(0)}), (\ref{r(infty)}),
(\ref{r(1)}) and the spectral decomposition formulae (\ref{solution-1})
, (\ref{solution-2}) and (\ref{solution-3}), one ends up with \cite{Del94b}
\begin{eqnarray}
&&\left \{\rho_\nu(x)\left (xq^{C(\nu)/2}
  +\epsilon(\nu)\epsilon(\nu') q^{C(\nu')/2}\right )-
  \rho_{\nu'}(x) \left (q^{C(\nu)/2}
  + \epsilon(\nu)\epsilon(\nu') x
  q^{C(\nu')/2}\right )\right \}\nonumber\\
&&~~~~~~~~~~\times~  {\cal P}^{\l\mu}_\nu\left (
  \pi_{\l}(e_0)\otimes \pi_{\mu}(q^{-h_0/2})\right )
  {\cal P}^{\l\mu}_{\nu'}=0
  \,,~~~\forall \nu\neq\nu'\,.
\label{ppp1}
\end{eqnarray}
If
\beq
{\cal P}^{\l\mu}_\nu\left (
  \pi_{\l}(e_0)\otimes \pi_{\mu}(q^{-h_0/2})\right )
  {\cal P}^{\l\mu}_{\nu'}\neq 0\label{pp2}
\eeq
then (\ref{ppp1}) gives rise to a relation between $\rho_\nu(x)$
and $\rho_{\nu'}(x)$,
\begin{equation}
\rho_\nu(x)=\rho_{\nu'}(x)\frac{q^{C(\nu)/2}
  + \epsilon(\nu)\epsilon(\nu') xq^{C(\nu')/2} }{xq^{C(\nu)/2}
  +\epsilon(\nu)\epsilon(\nu') q^{C(\nu')/2} }\;,~~~~
  \forall \nu\neq\nu'.\label{2rho1}
\end{equation}
It can be shown \cite{Del94b} that $\e(\nu)\e(\nu')=-1$ always if (\ref{pp2})
is satisfied. With the help of notation
\beq
\left\langle a\right\rangle\equiv \frac{1-xq^a}{x-q^a}
\eeq
(\ref{2rho1}) then becomes
\beq
\rho_\nu(x)=\left\langle\frac{C(\nu')-C(\nu)}{2}\right\rangle
  \rho_{\nu'}(x)\,.\label{2rho2}
\eeq

We have a relation between the coefficients
$\rho_\nu$ and $\rho_{\nu'}$ whenever the condition
\reff{pp2} is satisfied, i.e., whenever $\pi_\l(e_0)\otimes \pi_\mu(\qh)$
maps from the module $V(\nu')$ to the module $V(\nu)$.
As a graphical aid \cite{Zha91} we introduce the tensor product graph.

\begin{Definition}\label{reducibilitygraph}
The {\bf tensor product graph} $G^{\l\mu}$ associated to the tensor product
$V(\l)\otimes V(\mu)$ is a graph
whose vertices are the irreducible modules
$V(\nu)$ appearing in the decomposition \reff{free} of
$V(\l)\otimes V(\mu)$. There is an edge between a vertex $V(\nu)$
and a vertex $V(\nu')$ iff
\beq\label{link}
{\cal P}^{\l\mu}_\nu\left (\pi_\l(e_0)\otimes \pi_\mu(\qh)\right )
  {\cal P}^{\l\mu}_{\nu'}\neq 0.
\eeq
\end{Definition}

If $V(\l)$ and $V(\mu)$ are irreducible $U_q(\G)$-modules then the
tensor product graph is always connected, i.e., every node is linked to every
other node by a path of edges. This follows from the fact that
$\pi_\l(e_0)\otimes \pi_\mu(\qh)$ is related to the lowest component
of an adjoint tensor operator; for details see \cite{Zha91}.
This implies
that the relations \reff{2rho2} are sufficient to
determine all the coefficients $\rho_\nu(x)$ uniquely.
If the tensor product graph is multiply connected, i.e., if there exist
more than two paths between two nodes, then the relations overdetermine
the coefficients, i.e., there are consistency conditions.
However, because the existence of a solution to the Jimbo equations
is guaranteed by
the existence of the universal R-matrix, these consistency conditions
will always be satisfied.

The straightforward but tedious and impractical way to determine the tensor
product graph is to work out explicitly the left hand side  of (\ref{link}).
It is much more
practical to work instead with the following larger graph.

\begin{Definition}\label{tpg}
The {\bf extended tensor product graph} $\Gamma^{\l\mu}$ associated to the
tensor product $V(\l)\otimes V(\mu)$ is a graph
whose vertices are the irreducible modules
$V(\nu)$ appearing in the decomposition \reff{free} of
$V(\l)\otimes V(\mu)$. There is an edge between two vertices $V(\nu)$
and $V(\nu')$ iff
\beq
V(\nu')\subset V_{adj}\otimes V(\nu) ~~~\mbox{ and }~~
\epsilon(\nu)\epsilon(\nu')=-1.\label{adj}
\eeq
\end{Definition}
It follows again from the fact that $\pi_\l(e_0)\otimes \pi_\mu(\qh)$
is related to the lowest component of an adjoint tensor operator that
the condition in \reff{adj} is a necessary condition for \reff{link}
\cite{Zha91}.
This means that every link contained in the tensor product graph is contained
also in the extended tensor product graph but the latter may contain
more links. Only if the extended tensor product graph is a tree do we
know that it is equal to the tensor product graph. If we impose a
relation (\ref{2rho2}) on the $\rho$'s
for every link in the extended tensor product graph, we may be
imposing too many relations and thus may not always find a solution.
If however we do
find a solution, then this is the unique correct solution which we
would have obtained also from the tensor product graph.

The advantage of using the extended tensor product graph is that it can be
constructed using only Lie algebra representation theory.
We only need to be able to decompose tensor products and to determine
the parity of submodules.

\sect{New $R$-Matrices\label{rmatrix}\label{secexamples}}

We will now apply the technique developed in the last section to a
number of interesting examples.

\subsection{Examples of R-Matrices for $U_q(sl(m|n)^{(1)})$\label{slmn-e}}

As already seen in section \ref{secgl} we find it convenient to embedd the
root space of $sl(m|n)$ into the bigger space spanned by
$\{\e_i\}^m_{i=1}\bigcup \{\d_j\}^n_{j=1}$
and equipped with the non-definite inner product
\begin{equation}
(\e_i,\e_j)=\delta_{ij},~~~~
(\d_i,\d_j)=-\delta_{ij},~~~~ (\e_i,\d_j)=0,
\end{equation}
The root space is the subspace with no component in the direction of
$\sum_{i=1}^n\d_i$.
Any weight $\Lambda$ may be written as
\begin{equation}
\Lambda\equiv (\Lambda_1,\cdots,\Lambda_m|\bar{\Lambda}_1,
\cdots, \bar{\Lambda}_n)\equiv \sum_{i=1}^m\Lambda_i\e_i+
\sum_{j=1}^n\bar{\Lambda}_j\d_j.
\end{equation}
The graded half sum $\rho$ of the positive roots of $sl(m|n)$ is
\begin{equation}
2\rho=\sum_{i=1}^m(m-n-2i+1)\e_i+\sum_{j=1}^n(m+n-2j+1)\d_j\,.
\end{equation}
For convenience of notation we define the following elements
\beq
\l_b=\sum_{i=1}^b\e_i\,.
\eeq
Because an evaluation homomorphism exists for $U_q(sl(m|n)^{(1)})$
\cite{Zha92}\footnote{In \cite{Zha92} the author constructs
an evaluation homomorphism from $\overline{\uqgh}$ to $\uqg$. However,
one can show that
the evaluation map constructed in \cite{Zha92} also preserves the extra
$q$-Serre relations (\ref{e-S2}) and thus also provides a homomorphism
from $\uqgh$ to $\uqg$.\label{ev}},
every irrep of $U_q(sl(m|n))$ provides also an irrep for $U_q(sl(m|n)^{(1)})$.
Thus to any tensor product of two irreps of $U_q(sl(m|n))$ there corresponds
a spectral parameter dependent R-matrix.
Here, as examples, we will construct the R-matrices corresponding
to the following tensor products: (i) rank $a~(\geq 1)$ symmetric
tensor (which has the highest weight $a\l_1$) with
rank $b~(\geq 1)$ symmetric tensor of the same type;
(ii) rank $a$ antisymmetric tensor
with rank $b$ antisymmetric tensor of the same type.
Without loss of generality, we assume $m\geq a\geq b$ in the following.
For more information on the irreducible representations see Appendix
\ref{appclass}

We start with case (i). We assume the symmetric tensors to be contravariant.
The case of covariant symmetric tensors can be similarily treated. Such
contravariant symmetric tensors are obviously atypical and type (1) unitary.
In view of arguments in Appendix \ref{class} we
obtain the tensor product decomposition
\beq
V(a\l_1)\otimes V(b\l_1)=\bigoplus_c V(\L_c),
\eeq
where
\beq
\Lambda_c=(a+b-c)\e_1+c\e_2,~~~~~c=0,1,\cdots,b.
\eeq

The corresponding tensor product graph is
\beq
\unitlength=1mm
\linethickness{0.4pt}
\begin{picture}(114.60,10.60)(13,9)
\put(50.00,15.00){\circle*{5.20}}
\put(65.00,15.00){\circle*{5.20}}
\put(95.00,15.00){\circle*{5.20}}
\put(110.00,15.00){\circle*{5.20}}
\put(44.00,15.00){\makebox(0,0)[rc]{$V(a\l_1)\otimes V(b\l_1)~=$}}
\put(50.00,11.00){\makebox(0,0)[ct]{$\Lambda_0$}}
\put(65.00,11.00){\makebox(0,0)[ct]{$\Lambda_1$}}
\put(95.00,11.00){\makebox(0,0)[ct]{$\Lambda_{b-1}$}}
\put(110.00,11.00){\makebox(0,0)[ct]{$\Lambda_b$}}
\put(110.00,15.00){\line(-1,0){21.00}}
\put(50.00,15.00){\line(1,0){21.00}}
\put(80.00,15.00){\makebox(0,0)[cc]{$\cdots$}}
\end{picture}
\eeq
The Casimir takes the following values on the representations appearing in the
above graph
\beq
C(\Lambda_c)=(a+b-c)^2+c^2-2c+(a+b)(m-n-1).
\eeq
Using this information and \reff{2rho2} we can determine all the functions
$\rho_\nu(x)$ and we arrive at the spectral decomposition of the R-matrix
\beq
{\breve{R}}^{a\l_1,b\l_1}(x)=\rho_{\Lambda_0}(x)
\sum_{c=0}^b\prod_{i=1}^c\left\langle a+b-2i+2
  \right\rangle\,{\bf P}^{a\l_1,b\l_1}_{\Lambda_c}\label{glmnR1}
\eeq
(our convention here and below is that $\prod_{i=1}^0(\dots)=1$.).
The $a=b=1$ case had been worked out already in \cite{Baz88,Bra90}.
The overall scalar factor $\rho_{\Lambda_0}(x)$ is not
determined by the Jimbo equations or by the Yang-Baxter equation but
by the normalization condition \reff{unitary}.
We will from now on drop such overall
scalar factors from the formulas for the R-matrices.

It should be emphasized that although the spectral decomposition
(\ref{glmnR1}) of the R-matrices obtained from $\uqg$ formally
look like those for the R-matrices obtained from $U_q(sl(m+n))$,
they are actually completely different. For example
the ranks of the elementary intertwiners
${\bf P}^{a\l_1,b\l_1}_{\L_c}$ in (\ref{glmnR1}) differ from those in the case
of $U_q(sl(m+n))$ and thus the R-matrices (\ref{glmnR1}) have different
multiplicities of eigenvalues
from those for $U_q(sl(m+n))$. In particular, the R-matrices for
the case of $a=b=1$ are known to give rise to the Perk-Schultz model
R-matrices \cite{deV91,Mar92}. As an example,
let us compute the projectors for the $a=b=1$ case of $U_q(sl(2|1))$
explicitly. We have,
\bea
{\bf P}^{\l_1\l_1}_{\L_0}&=&[2]_q^{-1}\left ([2]_q(e_{11}
   +e_{55})+q(e_{22}+e_{33}
   +e_{66})+q^{-1}(e_{44}+e_{77}+e_{88})\right .\no\\
 & &\left . +e_{24}+e_{37}+e_{42}+e_{68}+e_{73}+e_{86}\right )\no\\
{\bf P}^{\l_1\l_1}_{\L_1}&=&1-{\bf P}^{\l_1\l_1}_{\L_0}
\eea
and the R-matrix is
\bea
\breve{R}^{\l_1\l_1}(x)&=&e_{11}+e_{55}+\frac{1-xq^2}{x-q^2}e_{99}+
  \frac{1-q^{-2}}{1-xq^{-2}}(e_{22}+e_{33}+e_{66}+x(e_{44}+e_{77}+e_{88}))\no\\
 & &~~~~~~~~~~+\frac{q^{-1}(1-x)}{1-xq^{-2}}(e_{24}+e_{37}+e_{68}
  +e_{42}+e_{73}+e_{86})\label{ps15}
\eea
where $e_{ij}$ is the matrix satisfying $(e_{ij})_{kl}=\d_{ik}\d_{jl}$.
(\ref{ps15}) is the Perk-Schultz 15-vertex model R-matrix.

For case (ii) we again assume the antisymmetric tensors to be contravariant.
Such contravariant antisymmetric tensors are atypical and type (1) unitary.
Following the arguments of Appendix \ref{class}, the tensor product
decomposition can be worked out to be
\beq
V(\l_a)\otimes V(\l_b)=\bigoplus_c V(\L_c)
\eeq
where, when $a+b\leq m$,
\beq
\Lambda_c=\sum_{i=1}^{a+c}\e_i
  +\sum_{i=1}^{b-c}\e_i,~~~~~c=0,1,\cdots,b
\eeq
and when $a+b>m$,
\bea
&&\Lambda_c=\sum_{i=1}^{a+c}\e_i
  +\sum_{i=1}^{b-c}\e_i,~~~~~c=0,1,\cdots,m-a\no\\
&&\Lambda_c=\sum_{i=1}^m\e_i+\sum_{i=1}^{b-c}\e_i+(a+c-m)\d_1,
  ~~~~~c=m-a+1,\cdots,b
\eea
The corresponding tensor product graph is
\beq
\unitlength=1mm
\linethickness{0.4pt}
\begin{picture}(112.60,7.60)(10,12)
\put(50.00,15.00){\circle*{5.20}}
\put(65.00,15.00){\circle*{5.20}}
\put(95.00,15.00){\circle*{5.20}}
\put(110.00,15.00){\circle*{5.20}}
\put(44.00,15.00){\makebox(0,0)[rc]{$V(\l_a)\otimes V(\l_b)~=$}}
\put(50.00,11.00){\makebox(0,0)[ct]{$\Lambda_0$}}
\put(65.00,11.00){\makebox(0,0)[ct]{$\Lambda_1$}}
\put(95.00,11.00){\makebox(0,0)[ct]{$\Lambda_{b-1}$}}
\put(110.00,11.00){\makebox(0,0)[ct]{$\Lambda_b$}}
\put(110.00,15.00){\line(-1,0){21.00}}
\put(50.00,15.00){\line(1,0){21.00}}
\put(80.00,15.00){\makebox(0,0)[cc]{$\cdots$}}
\end{picture}
\eeq
The Casimirs are
\beq
C(\Lambda_c)=2c(b-a-1)-2c^2+(m-n+1)(a+b)+2b-a^2-b^2
\eeq
and we obtain the following spectral decomposition of the R-matrix
\beq
{\breve{R}}^{\l_a,\l_b}(x)=\sum_{c=0}^b\prod_{i=1}^c\left\langle
  2i+a-b\right\rangle\,{\bf P}
 ^{\l_a,\l_b}_{\Lambda_c}
\eeq

\subsection{R-Matrices for $U_q(sl(1|n)^{(1)})$ with Extra Continuous
Parameters\label{secpar1}}

It is well known \cite{Kac78} that type-I superalgebras admit nontrivial
one-parameter families of finite-dimensional irreps which deform to provide
one-parameter families of finite-dimensional irreps of the corresponding
type-I quantum superalgebras \cite{Gou93}.

For an indeterminate $\a\in {\bf R}$ we will consider
the one-parameter family of $2^n$-dimensional irreducible
$U_q(sl(1|n))$-modules $V(\alpha)$ with highest weights of the form
$\Lambda(\alpha)=(\a|0,\cdots,0)\equiv \a\e$. From the classification scheme
given in Appendix \ref{class}, $V(\a)$ is a unitary irreducible module provided
that $0<q\in {\bf R}$ and $\a>n-1$ or $\a<0$. It follows that the
tensor product module $V(\a)\otimes V(\b)$ is completely reducible for
$\a,~\b>n-1$ or $\a,~\b<0$. The decomposition is given by
\beq
V(\a)\otimes V(\b)=\bigoplus_{c=0}^n V(\L_c)\label{decom1}
\eeq
with
\beq
\L_c=(\a+\b-c)\e+\l_c,~~~~~
\l_c=\sum_{i=1}^c\d_i
\eeq
To see this, we observe that each highest weight occuring in the decomposition
is of the form of $\a\e+\mu$, where $\mu$ is an element of the weight
spectrum of $V(\b)$. Moreover, to ensure that $\a\e+\mu$ is a dominant
weight for $sl(1|n)$, the weight $\mu$ must be dominant for
$u(1)\bigoplus sl(n)$. From the induced module construction of Appendix
\ref{induced}, we see that $\mu$ must be of the form
$\mu_c=(\b-c)\e+\l_c$ and it occurs with unit multiplicity. It is easily
checked by dimensions that all highest weights $\L_c=\a\e+\mu_c$ arise in
the decomposition, leading to (\ref{decom1}).

The tensor product graph in this case is simply
\beq
\unitlength=1mm
\linethickness{0.4pt}
\begin{picture}(112.60,7.60)(10,12)
\put(50.00,15.00){\circle*{5.20}}
\put(65.00,15.00){\circle*{5.20}}
\put(95.00,15.00){\circle*{5.20}}
\put(110.00,15.00){\circle*{5.20}}
\put(44.00,15.00){\makebox(0,0)[rc]{$V(\a)\otimes V(\b)~=$}}
\put(50.00,11.00){\makebox(0,0)[ct]{$\Lambda_0$}}
\put(65.00,11.00){\makebox(0,0)[ct]{$\Lambda_1$}}
\put(95.00,11.00){\makebox(0,0)[ct]{$\Lambda_{n-1}$}}
\put(110.00,11.00){\makebox(0,0)[ct]{$\Lambda_n$}}
\put(110.00,15.00){\line(-1,0){21.00}}
\put(50.00,15.00){\line(1,0){21.00}}
\put(80.00,15.00){\makebox(0,0)[cc]{$\cdots$}}
\end{picture}
\eeq
Using the fact that $\L_{c+1}=\L_c-\e+\d_{c+1}$, we obtain the differences
of the Casimirs
\beq
\frac{C(\L_c)-C(\L_{c+1})}{2}=\a+\b-2c.
\eeq
For the R-matrix we find
\beq
\breve{R}^{\a\b}(x)=\sum_{c=0}^n\prod_{i=1}^c\left\langle \a+\b-2i+2
  \right\rangle {\bf P}^{\a\b}_{\L_c}.
\eeq
This R-matrix is obtained for $0<q\in {\rm R}$ and $\a,\b>n-1$ or
$\a,\b<0$ but should hold for other values of $q,~\a,~\b$ via analytic
continuation arguments.

It is worth emphasizing that $\breve{R}^{\a\b}(x)$ satisfies the QYBE
\beq
(I\otimes\breve{R}^{\a\b}(x))(\breve{R}^{\a\g}(xy)\otimes I)(I\otimes
   \breve{R}^{\b\g}(y))=(\breve{R}^{\b\g}(y)\otimes I)(I\otimes
   \breve{R}^{\a\g}(xy)\otimes I)(\breve{R}^{\a\b}(x)\otimes I)\label{non-a1}
\eeq
acting on $V(\a)\otimes V(\b)\otimes V(\g)$ and the parameters $\a,~\b$
and $\g$ enter the QYBE in a non-additive form.

We should remark that although it appears that the eigenvalues of
$\breve{R}^{\a\b}(x)$ depend only on $\a+\b$, the elementary intertwiners
${\bf P}^{\a\b}_{\L_c}$ depend on both $\a$ and $\b$. For example, for
$U_q(sl(1|1))$~\footnote{
$sl(1|1)$ is not a simple algebra, but our method give
nevertheless a valid R-matrix, as the example shows.},
the ${\bf P}^{\a\b}_{\L_0}$ and ${\bf P}^{\a\b}_{\L_1}$
in the above equation take the form
\begin{eqnarray}
{\bf P}^{\alpha\beta}_{\Lambda_0}&=& [\alpha+\beta]_q^{-1}\;\left (
\begin{array}{cccc}
[\alpha+\beta]_q & 0 & 0 & 0\\
0 & ([\alpha]_q[\beta]_q)^{1/2}\,q^{(\alpha+\beta)/2} & [\alpha]_q & 0 \\
0 & [\beta]_q & ([\alpha]_q[\beta]_q)^{1/2}\,q^{-(\alpha+\beta)/2} & 0 \\
0 & 0 & 0 & 0
\end{array}
\right )\,,\nonumber\\
{\bf P}^{\alpha\beta}_{\Lambda_1}&=& [\alpha+\beta]_q^{-1}\;\left (
\begin{array}{cccc}
0 & 0 & 0 & 0\\
0 & ([\alpha]_q[\beta]_q)^{1/2}\,q^{-(\alpha+\beta)/2} & -[\beta]_q & 0 \\
0 & -[\alpha]_q & ([\alpha]_q[\beta]_q)^{1/2}\,q^{(\alpha+\beta)/2} & 0 \\
0 & 0 & 0 & [\alpha+\beta]_q
\end{array}
\right )\,.\label{pp1}
\end{eqnarray}
and the R-matrix in this case reads
\begin{equation}\label{R}
\breve{R}^{\alpha\beta}(x)=\left (
\begin{array}{cccc}
1 & 0 & 0 & 0\\
0 & -([\alpha]_q[\beta]_q)^{1/2}q^{(\alpha+\beta)/2}\cdot \frac{q-q^{-1}}
 {x-q^{\alpha+\beta}} & \frac{xq^\beta-q^\alpha}{x-q^{\alpha+\beta}} & 0\\
0 & \frac{xq^\alpha-q^\beta}{x-q^{\alpha+\beta}} &
 -([\alpha]_q[\beta]_q)^{1/2}q^{(\alpha+\beta)/2}\cdot \frac{x(q-q^{-1})}
 {x-q^{\alpha+\beta}} & 0\\
0 & 0 & 0 & \frac{1-xq^{\alpha+\beta}}
  {x-q^{\alpha+\beta}}
\end{array}
\right )\label{gl11-R}
\end{equation}
More details can be found in \cite{Bra94b}. It should be pointed out that
(\ref{gl11-R}) is the trigonometric limit, up to a similarity transformation
(c.f. \cite{Bra93}), of the 8-vertex free fermion model R-matrix with extra
non-additive parameters, obtained in
\cite{Baz85}.

\sect{Discussion\label{secdiscussion}}

By extending the formalism of \cite{Del94b} for quantum bosonic algebras, a
systematic method was developed for obtaining trigonometric solutions
$\breve{R}(x)\in {\rm End}(V(\l)\otimes V(\mu))$ to the QYBE for the
type-I quantum superalgebras $U_q(\G)$, where $V(\l),~V(\mu)$ are irreducible
$U_q(\G)$-modules such that their tensor product is completely reducible
and multiplicity-free. In this connection, it is worth noting that the
type-I quantum superalgebras admit two distinct (large) classes of
unitary irreps, classified in refs.\cite{Gou93,Lin93} and discussed in
Appendix \ref{class}, so that $V(\l)\otimes V(\mu)$ is automatically
completely reducible provided $V(\l),~V(\mu)$ are both unitary of the
same type.

Our approach, which is based on the tensor product graph method, enables the
construction of a large number of new R-matrices. As explicit examples,
we have considered the cases where $V(\l),~V(\mu)$ correspond to the
contravariant symmetric or anti-symmetric tensor irreps of $\uqg$ in
section \ref{slmn-e}. They give rise to generalizations of the
Perk-Schultz model R-matrices.

As noted in the introduction, quantum superalgebras are not straightforward
generalizations of quantum algebras but have a far richer structure and
representation theory. In particular, the type-I quantum superalgebras
have the intriguing property that they admit one-parameter families of
irreps $V(\L(\a)),~\L(\a)\equiv \L+\a\sum_i\e_i$, corresponding to each
typical $\L\in D_+$ (c.f. Appendix \ref{appinduced}). For real $\a$
sufficiently large these irreps are all unitary, typical and have the same
dimension. This enables the construction of new R-matrices
$\breve{R}^{\a\b}(x)\in {\rm End} [V(\L(\a))\otimes V(\L(\b))]$ depending
on two extra parameters $\a,~\b$.

It should be emphasized that these R-matrices do not satisfy the usual
QYBE (except when $\a=\b$) but rather its natural extension,
eq.(\ref{non-a1}), in which the parameters $\a,~\b$ enter in a way
analogous to the spectral parameter $x$ but in a non-additive form.
Nevertheless, as will be shown elsewhere, solutions to this equation give
rise to exactly solvable models and the entire apparatus of the QISM,
including the (nested) algebraic Bethe ansatz, can be extended for the
treatment of these models. In this way we obtain new three parameter
exactly solvable models generalizing the (trigonometric) free fermion
model \cite{Baz85}. The latter in fact arises from a certain one
parameter family of two dimensional irreps of $U_q(sl(1|1))$, as we
have shown in the paper.

Further new examples of such R-matrices arising from one-parameter families
of irreps for $U_q(sl(1|n))$ (which includes the free fermion six vertex
model when $n=1$) were obtained in section \ref{secpar1}
. In future work we aim to investigate
the exactly solvable models arising from these trigonometric R-matrices
(and their rational limits) and the physical significance of the two
extra parameters, $\a,~\b$.

\begin{center}
{\bf Acknowledgements:}
\end{center}
We thank Anthony J. Bracken and Rui Bin Zhang for discussions. The financial
support from the Australian Research Council is gratefully acknowledged.

\newpage

\appendix

\sect{Kac's Induced Module Construction\label{induced}\label{appinduced}}

For self-containedness, in this appendix we will give a brief overview of
the induced module construction for type-I superalgebras
due to Kac \cite{Kac77,Kac78}. However, as is shown in \cite{Zha93a,Zha93b},
the construction also applies in the quantum case for generic $q$,
giving modules with the same dimension and weight spectrum as the $q=1$ case.

Let $\G$ denote a Lie superalgebra with the distinguished ${\bf Z}$
gradation \cite{Kac77},
\beq
\G=\bigoplus_{i\in{\bf Z}}{\G}_i\label{z-gradation}
\eeq
satisfying $[{\G}_i,{\G}_j]={\G}_{i+j}$.
For type-I Lie superalgebras, (\ref{z-gradation}) simplifies to
$\G={\G}_{-1}\bigoplus {\G}_0\bigoplus {\G}_1$.
This immediately implies that for $x,y\in{\G}_{-1},~$
$[x,y]=xy+yx=0~\Longrightarrow~xy=-yx$,
and in particular, $x^2=0$.

Let $\D_1^+$ denote the positive odd roots of $\G$. We have that
$\{F_\a\,|\,\a\in\D^+_1\}$ form a basis for $\G_{-1}$ and
$\{E_\a\,|\,\a\in\D^+_1\}$ a basis for $\G_{1}$. We refer to $\G_0$ as the
``even subalgebra" of $\G$. Specifically, we have
$\G_0=u(1)\bigoplus sl(m)\bigoplus sl(n)\,,~~{\rm for}~\G=sl(m|n)$ with
$m,n\geq2$,~
$\G_0=u(1)\bigoplus sl(n)\,,~~{\rm for}~\G=sl(1|n),
{}~n\geq2$ and
$\G_0=u(1)\bigoplus sp(2n)\,,~~{\rm for}~\G=osp(2|2n)$.
{}From (\ref{z-gradation}) we see that
$[\G_0,\G_{-1}]=\G_{-1}$
so that $\G_{-1}$ provides a representation of $\G_0$. Let $U(\G_{-1})$
denote the universal enveloping algebra of $\G_{-1}$.
{}~$U(\G_{-1})$ is finite dimensional and the P.B.W
theorem takes a particularly simple form. We choose the following
basis elements for $U(\G_{-1})$
\beq
\left \{ \Gamma(\hat{t})=\prod_{\a\in\D^+_1}(F_\a)^{t_\a},~~~~t_\a=0,1\rt\}
\eeq
which are unique up to a sign. These elements generate a finite dimensional
module for $\G_0$ under the adjoint action.

Now let $V_0(\L)$ denote an irreducible $\G_0$-module  and
set $\overline{V(\L)}$ to be the tensor product
module
\beq
\overline{V(\L)}=U(\G_{-1})\otimes_{U(\G_0)}\, V_0(\L)\,.
\eeq
We refer to $\overline{V(\L)}$ as the Kac module.

The $\G$-module $\overline{V(\L)}$ is not necessarily irreducible. If
it is, we set
$V(\L)=\overline{V(\L)}$
and refer to $\L$ and $V(\L)$ as ``typical". On the other hand, if
$\overline{V(\L)}$ contains a proper maximal submodule $M$ (necessarily
unique) we set
$V(\L)=\overline{V(\L)}\,/\,M$
and refer to $\L$ and $V(\L)$ as ``atypical". There is a well known
criterion for typicality of $V(\L)$ due to Kac \cite{Kac78}, that is
$V(\L)$ is typical $\G$-module iff $(\L+\rho,\a)\neq 0\,,~\forall \a
\in\D^+_1$.

Let us remark that for typical modules the dimensions are easily evaluated
to be ${\rm dim}V(\L)=2^d\cdot{\rm dim}V_0(\L)$,
where $d$ (which equals to $mn$ for $sl(m|n)$ and to $2n$ for $osp(2|2n)$)
is the number of odd positive roots. This formula is
particularly useful in determining tensor product decompositions of typical
modules.

The induced module construction provides
an insight into the existence of one-parameter families of irreps. Observe
that for the type-I superalgebras, the even subalgebra $\G_0$ contains a
$u(1)$ subalgebra. Hence when we take the $\G_0$-module $V_0(\L)$, the
action of the $u(1)$ subalgebra is simply scalar multiplication by an
arbitrary $\a\in {\bf\rm C}$. When we take the Kac module, we thereby obtain a
one-parameter family of modules with the same dimension.


\sect{Classification Theorems of Finite-Dimensional Irreps\label{class}
\label{appclass}}

The type-I superalgebras admit two types of unitary representations
which may be described as follows. We define a conjugation operation on
$\G$ generators by
$e_i^\dagger=f_i\,,~~~~f_i^\dagger=e_i\,,~~~~h_i^\dagger=h_i$
which is extended uniquely to all of $\G$ such that
$(uv)^\dagger=v^\dagger u^\dagger\,,~~~~~~\forall u,v\in\G$.
If $\pi_\L$ denotes a representation of highest weight $\L$ then we call
$\pi_\L$ type (1) unitary if
\beq
\pi_\L(u^\dagger)=\overline{\pi_\L(u)}\,,~~~~\forall u\in \G
\eeq
and type (2) unitary if
\beq
\pi_\L(u^\dagger)=(-1)^{[u]}\;\overline{\pi_\L(u)}\,,~~~~
  \forall u\in \G
\eeq
where the overline denotes Hermitian matrix conjugation.
The two types of unitary representations are in fact related via duality.

Such unitary representations have the property that they are always
completely reducible and the tensor product of two representations of the
same type reduce completely into unitary representations of the same type.

The finite dimensional irreducible unitary representations have been
classified in \cite{Gou90}. Reference \cite{Gou90} deals with
the irreps of $gl(m|n)$ rather than $sl(m|n)$. There is however
no substantial difference besides the fact that some irreps which are
distinct as irreps of $gl(m|n)$ are isomorphic as irreps of
$sl(m|n)$.

\begin{Theorem}\label{theo1}
A given $gl(m|n)$-module
$V(\L)$, with $\L\in D_+$, is type (1) unitary iff: i) $(\L+\rho,\e_m
-\d_n)>0$; or ii) there exists an odd index $\mu\in\{1,2,\cdots,n\}$
such that $(\L+\rho,\e_m-\d_\mu)=0=(\L,\d_\mu-\d_n)$.
In the former case the given condition also enforces typicality on
$V(\L)$, while in the latter case all irreps are atypical.
\end{Theorem}

\begin{Theorem} \label{theo2}
A given $gl(m|n)$-module
$V(\L)$, with $\L\in D_+$, is type (2) unitary iff: i) $(\L+\rho,\e_1
-\d_1)<0$; or ii) there exists an even index $k\in\{1,2,\cdots,m\}$
such that $(\L+\rho,\e_k-\d_1)=0=(\L,\e_1-\e_k)$.
In the former case $V(\L)$ is typical, while in the latter case it is atypical.
\end{Theorem}

The so-called contravariant and covariant tensor irreps for
$gl(m|n)$ were studied using the Young super-diagram method
in \cite{Don81,Bal81}. From \cite{Gou90}, we have

\begin{Proposition}\label{prop1}
The contravariant and covariant
tensor irreps of $gl(m|n)$ are unitary irreps of type (1) and type (2),
respectively.
\end{Proposition}

Except for the typical type (1) unitary irreps with $c=(\L+\rho,\e_m-
\d_n)>0$ being {\em non-integral}, the rest of the type (1) unitary
irreps include the contravariant tensor irreps.
Similarily the atypical type (2) unitary irreps and the typicals with
$c'=(\L+\rho,\e_1-\d_1)$ being an integer, include the
covariant tensor irreps of $gl(m|n)$.

Consider a contravariant tensor
irrep of $gl(m|n)$ which is characterized by a partition of the Young
super-diagram,
\beq
P=(p_1,p_2,\cdots,p_m,p_{m+1},p_{m+2},\cdots,p_N)
\eeq
where the box numbers $p_a\in {\bf Z}^+\,,~p_a\geq p_{a+1},~\forall a$
and $p_{m+1}\leq n$ is assumed. Associated with each partition $P$, there
exists a unique highest weight
\beq
\L=(\L_1,\L_2,\cdots,\L_m|\bar{\L}_1,\bar{\L}_2,\cdots,\bar{\L}_n)
\eeq
with
\beq
\L_i=p_i\,,~~~~~~i=1,2,\cdots,m
\eeq
and the $\bar{\L}_i$'s defined by
\beq
\sum_{i=1}^n\bar{\L}_i\d_i=\sum_{i_1=1}^{p_{m+1}}\d_{i_1}
    +\sum_{i_2=1}^{p_{m+2}}\d_{i_2}+\cdots +
    \sum_{i_{N-m}=1}^{p_N}\d_{i_{N-m}}
\eeq
When $p_m\geq n$, the irreducible $gl(m|n)$-module associated with
the partition $P$ is typical and type (1) unitary. If $p_m< n$, it is atypical
and type (1) unitary.


The reduction of tensor products in the non-graded case extendeds to the
graded case \cite{Don81,Bal81}: the Kronecker products for the same type of
irreps of $gl(m|n)$ (i.e. contravariant with contravariant, or covariant with
covariant, but not a mixture of them) are governed by the
usual (Littlewood-Richardson) rule. However, there is a major difference:
for $gl(k)$ the Young tableaux can have at most $k-1$ rows but for $gl(m|n)$
the Young super-tableaux can have any number of rows. This is because for
superalgebras ``antisymmetrization" of basis states implies antisymmetrization
of bosonic parts but symmetrization of fermionic parts and one can
continue the ``antisymmetrization" process up to infinity.

We should emphasize that when $c$ (resp. $c'$) is non-integral,
there exists non-trivial one-parameter
family of typical type (1) (resp. type (2)) unitary irreps, which
are not tensorial. Thus those irreps cannot be dealt with by the Young
super-diagram techniques.
In particular, for any one-parameter family of Kac modules of $gl(m|n)$ with
highest weights of the form
$\L(\a)\equiv \L+\a\sum_{i=1}^m\e_i$,~
$\pi_{\L(\a)}$ is both unitary and typical for $|\a|$ sufficiently large.
We then know that $V(\L(\a))\otimes V(\L(\b))$ is completely reducible.
This provides an initial step in calculating the tensor product
decomposition in previous sections.

As shown in \cite{Gou93,Lin93,Zha92,Zha93a,Zha93b,Lin94}, all of the above
considerations apply equally well to the type-I quantum superalgebra
$U_q(sl(m|n))$.
In particular, the classification theorems still hold provided that
$\L$ is real and $q>0$.


\end{document}